\documentclass[aip,reprint]{revtex4-2}
\usepackage{amsfonts,amsmath,bm,amssymb,color}
\usepackage{graphicx}
\usepackage[makeroom]{cancel}

\renewcommand{\Re}{\mathrm{Re}}

\newcommand{\Ro}{\operatorname{Ro}}

\usepackage{rotating}
\usepackage{tikz,tkz-euclide}
\usetikzlibrary{decorations.pathmorphing}
\usetikzlibrary{arrows}
\usetikzlibrary{arrows.meta}
\usetikzlibrary{decorations.markings}
\usepackage{pgfplots}
\usepackage{tkz-fct}
\usetikzlibrary{calc,patterns}

\usepackage[unicode=true]{hyperref}

\definecolor{bostonuniversityred}{rgb}{0.8, 0.0, 0.0}
\definecolor{dukeblue}{rgb}{0.0, 0.0, 0.61}
\definecolor{ao(english)}{rgb}{0.0, 0.5, 0.0}
\definecolor{darkmagenta}{rgb}{0.55, 0.0, 0.55}
\definecolor{armygreen}{rgb}{0.29, 0.33, 0.13}
\definecolor{coquelicot}{rgb}{1.0, 0.22, 0.0}
\definecolor{fucsiak}{rgb}{0.4, 0.08, 0.4}
\definecolor{airforceblue}{rgb}{0.36, 0.54, 0.66}

\begin{document}

\title{Velocity profiles of cyclones and anticyclones in a rotating turbulent flow}
\author{Vladimir M. Parfenyev}
\email{parfenius@gmail.com}
\author{Ivan A. Vointsev}
\author{Alyona O. Skoba}
\author{Sergey S. Vergeles}
\email{ssver@itp.ac.ru}

\affiliation{Landau Institute for Theoretical Physics, Russian Academy of Sciences,\\1-A Akademika Semenova av., 142432 Chernogolovka, Russia}
\affiliation{National Research University Higher School of Economics, Faculty of Physics,\\Myasnitskaya 20, 101000 Moscow, Russia}

\begin{abstract}
Strong rotation makes an underlying turbulent flow quasi-two-dimensional that leads to the upscale energy transfer. Recent numerical simulations show that under certain conditions, the energy is accumulated at the largest scales of the system, forming coherent vortex structures known as condensates. We analytically describe the interaction of a strong condensate with weak small-scale turbulent pulsations and obtain an equation that allows us to determine the radial velocity profile $U(r)$ of a coherent vortex. When external rotation is fast, the velocity profiles of cyclones and anticyclones are identical to each other and are well described by the dependence $U(r) \propto \pm r \ln (R/r)$, where $R$ is the transverse size of the vortex. As the external rotation decreases, this symmetry disappears: the maximum velocity in cyclones is greater and the position of the maximum is closer to the axis of the vortex in comparison with anticyclones. Besides, our analysis shows that the size $R$ of the anticyclone cannot exceed a certain critical value, which depends on the Rossby and Reynolds numbers. The maximum size of the cyclones is limited only by the system size under the same conditions. Our predictions are based on the linear evolution of turbulent pulsations on the background of the coherent vortex flow and are accompanied by estimates following from the nonlinear Navier-Stokes equation.
\end{abstract}

\maketitle

\section{Introduction}

It is well known that rotation plays a crucial role in determining the properties of underlying turbulence in three-dimensional flows, see, e.g., the recent review~\cite{godeferd2015structure}. When rotation is strong, the velocity field tends to become quasi-two-dimensional varying weakly along the rotation axis following the Taylor-Proudman theorem~\cite{proudman1916motion, taylor1917motion}. For this reason, it is often said that strong rotation effectively makes the flow two-dimensional and opens the way for an inverse energy cascade (from small to large scales), which is characteristic of two-dimensional turbulence~\cite{boffetta2012two}. If the damping of the large-scale motion in the system is small enough, then the inverse cascade can lead to the formation of large-scale columnar vortices that are stable over many periods of their rotation (the so-called condensate).

The spontaneous formation of long-living columnar vortices aligned in the direction of rotation in three-dimensional systems has been observed both in experiments~\cite{mcewan1976angular, hopfinger1982turbulence, ruppert2005extraction, staplehurst2008structure, moisy2011decay, gallet2014scale, boffetta2020cyclone} and in numerical simulations~\cite{bartello1994coherent, yeung1998numerical, smith1999transfer, yoshimatsu2011columnar, biferale2016coherent, seshasayanan2018condensates}. Remarkably, the distribution of these vortices is asymmetric with the predominance of cyclones (co-rotating with the external rotation) over anticyclones. The qualitative explanation is that anticyclones weaken the external rotation due to their own rotation which prevents the development of the inverse energy cascade inside them. Quantitatively, the cyclone-anticyclone asymmetry is studied by statistical methods using various correlation functions~\cite{gallet2014scale, deusebio2014dimensional, naso2015cyclone} that involve spatial averaging over the entire observation area.

This approach is justified, since in most of the works mentioned above, the large-scale flow is not strictly stationary and at large times required for collecting statistics, the configuration of cyclones and anticyclones changes. The only exception among the experiments known to us was reported in Ref.~\onlinecite{ruppert2005extraction}, where truly stationary vortices were observed and their spatial arrangement was fixed by inhomogeneity in the permanent forcing of the flow. However, the results of recent numerical simulations demonstrate that with increasing external rotation, columnar vortices become very long-lived~\cite{seshasayanan2018condensates}. Besides, their transverse size increases; it exceeds the forcing scale and becomes comparable to the system size $L$. The velocity amplitude $U\sim L \sqrt{\epsilon/\nu}$ inside the vortex is the result of a balance between the forcing power $\epsilon$ per unit mass and the viscous friction inside the condensate, where $\nu$ is the kinematic viscosity of a fluid. In this state, new questions arise related to the structure of an individual vortex. How does the azimuthal velocity of the vortex change depending on the distance to its axis? How does the vortex velocity profile depend on the global parameters of the problem? Is there any difference between cyclones and anticyclones?

The first attempt to address these questions was made in Ref.~\onlinecite{kolokolov2020structure}, where the authors studied the limiting case of fast rotation (Rossby number $\mathrm{Ro}_{\scriptscriptstyle R} =\sqrt{\epsilon/\nu}/(2\Omega_0)\ll 1$, where $\Omega_0$ is the angular velocity of the external rotation) and small-scale forcing ($k_f L\gg 1/\mathrm{Ro}_{\scriptscriptstyle R}$, where $k_f$ is the characteristic wavenumber of the forcing). In this limiting case, the power balance between the forcing and the viscous friction is local in space. The forcing first excites small-scale inertial waves and then their kinetic energy is transferred to the large-scale vortex flow. This leads to the universal velocity profile $U(r) \sim  \pm \sqrt{\epsilon/\nu} \,r \ln(R/r)$, which describes both cyclones and anticyclones. Here $r$ is the distance to the vortex axis and $R\lesssim L$ is the transverse vortex size. The absence of cyclone-anticyclone asymmetry in the limit of small Rossby numbers is in agreement with the results of numerical simulations~\cite{seshasayanan2018condensates}. Note that the local power balance between the forcing and the viscous dissipation inside the coherent flow can be achieved in two-dimensional flow as well, resulting in the same linear-logarithmic velocity profile observed in Ref.~\onlinecite{doludenko2021coherent}. As for experimental attempts to measure the velocity profiles of coherent vortices, we would like to mention works~\onlinecite{mcewan1976angular, hopfinger1982turbulence}, although the condensates in them were clearly underdeveloped (large-scale vortices live for a relatively short time and their size is small compared to the system size).

Here we consistently study the structure of the vortex condensates assuming that the Rossby number is moderate and analyze the cyclone-anticyclone asymmetry in the developed condensate. For this purpose, we extend the approach proposed in Ref.~\onlinecite{kolokolov2020structure} to the case of moderately low Rossby numbers. We consider the turbulent three-dimensional pulsations on the background of the coherent vortex flow in the linear regime, assuming that the rapid distortion theory is applicable. The turbulent pulsations produce the Reynolds stress which maintains the coherent vortex. Then we obtain the mean velocity profile $U(r)$ and analyze the difference between cyclones and anticyclones. Using the estimates following from the full nonlinear Navier-Stokes equation, we derive the criteria when the coherent vortices can exist.

Actually, the cyclone-anticyclone asymmetry can be observed even at a small Rossby number if the system size is large enough. Inside a coherent vortex, the external rotation is superimposed by self-rotation $U(r)/r \sim \pm \sqrt{\epsilon/\nu} \ln(R/r)$ induced by the vortex. The total local rotation rate is equal to $\Omega = \Omega_0 + U(r)/r$, so the local Rossby number $\mathrm{Ro}(r) = r \partial_r (U/r)/(2 \Omega)$ formally grows indefinitely with approaching the axis of anticyclone. However, the linear-logarithmic profile is not applicable below some core radius $r_c$, where the local power balance approximation is no longer valid and the profile changes to a rigid body rotation~\cite{kolokolov2020structure}. Thus, one can estimate the maximum possible radius of an anticyclone from the condition $2\ln(R_{max}/r_c)\sim 1/\mathrm{Ro}_{\scriptscriptstyle R}$, which ensures that the local Rossby number inside the anticyclone core remains below unity to prevent the development of direct energy cascade (characteristic of a three-dimensional turbulent flow with weak rotation). The restriction $R<R_{max}$ for anticyclones means that the cyclone-anticyclone asymmetry occurs in the case $L \gg R_{\max}$, since in cyclones the self-rotation increases the total rotation rate, so there is no similar limitation for them.

\section{Momentum balance}

We consider a three-dimensional incompressible turbulent fluid that rotates around the $Z$-axis with a constant angular velocity $\bm{\Omega}_0$. In the rotating frame, the system is described by the Navier-Stokes equation with an additional term describing the Coriolis force and with an external force $\bm{f}$ random in time and space with zero mean (pumping or forcing), which continuously excites small-scale turbulent pulsations
\begin{equation}\label{ns_init}
    \partial_t\bm v+(\bm v \nabla)\bm v+2\left[\bm\Omega_0,\bm v\right]=-\nabla p+\nu\Delta\bm v + \bm{f}.
\end{equation}
Equation (\ref{ns_init}) has to be supplemented by the incompressibility condition $\mathrm{div} \, \bm v = 0$. Here $\bm v$ is the fluid velocity, $p$ is the effective pressure divided by the fluid mass density, which differs from the physical pressure by addition of the potentials produced by the centrifugal and gravitational forces, and $\nu$ is the kinematic viscosity coefficient. We assume that the force $\bm f$ produces the energy flux per unit mass $\epsilon$, has a correlation length in space equal to $1/k_f \ll L$, where $L$ is the system size, and is shortly correlated in time. The Reynolds number characterizing the pumping should be large, ${\mathrm{Re}}_f = \epsilon^{1/3}/\nu k_f^{4/3}\gg1$, which is a prerequisite for the overall flow to be turbulent.

\begin{figure}
\centering{\includegraphics[width=0.8\linewidth]{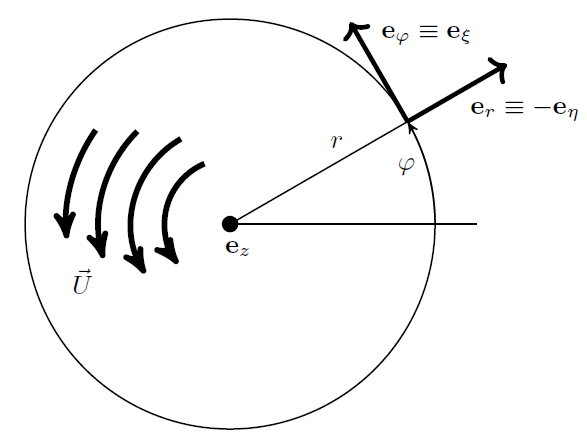}}
\caption{Schematic of a coherent columnar vortex and the introduced cylindrical $\{{r},\varphi,z\}$ and Cartesian $\{\xi,\eta,z\}$ coordinate systems.}
\label{fig:vortexr}
\end{figure}

Next, we assume that a coherent columnar large-scale vortex has already formed in our system. We introduce a cylindrical coordinate system $\{{r},\varphi,z\}$ so that the vortex axis coincides with the direction of the $Z$-axis, see Fig.~\ref{fig:vortexr}. In the vortex flow region, the velocity field can be represented as the sum of the mean flow $\bm U$ and turbulent pulsations $\bm u$:
\begin{equation}
\bm v= \bm u+\bm U,\quad \bm u=\{u_{{r}}, u_{\varphi}, u_z\},\quad \bm U=\{0, U, 0\},
\label{u_and_U}
\end{equation}
where, by definition, the average of the fluctuations is equal to zero, $\langle \bm u \rangle=0$. Substituting this representation into equation (\ref{ns_init}), averaging the result over turbulent pulsations, and projecting the equation onto ${\bf e}_\varphi$, we obtain
\begin{equation}
\partial_t U = -\left( \partial_{{r}} + \frac{2}{{r}} \right) \Pi^{\varphi{r}},
\quad
\Pi^{\varphi{r}} = \langle u^{\varphi} u^{{r}} \rangle - \nu \Sigma,
\label{eq_U}
\end{equation}
where $\Pi^{ij}$ is the Reynolds stress tensor collected with the viscous term. Here we also introduced the local large-scale shear rate $\Sigma = {r} \, \partial_{{r}}(U/{r})$ which characterizes the difference between local rotation in the vortex and rigid body rotation.

\begin{figure*}
\centering{\includegraphics[width=0.7\linewidth]{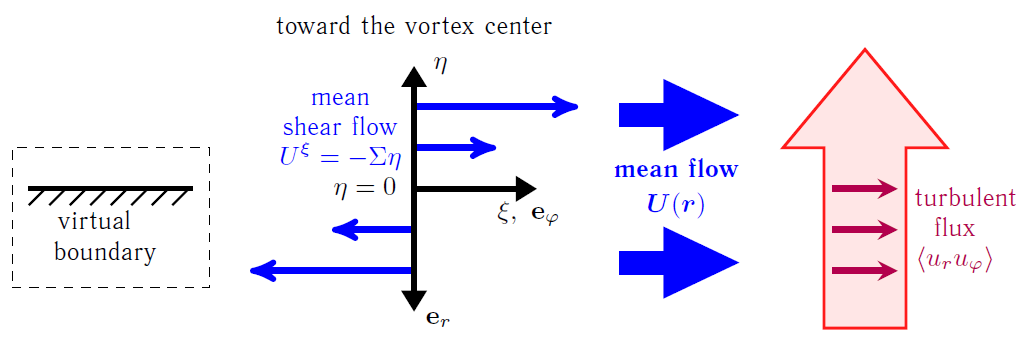}}
\caption{Reynolds stress tensor, for definiteness $\Sigma(r)<0$ is chosen. The red arrow depicts the turbulent part of the $\xi$-component of the momentum flux in $\eta$-direction. Virtual boundary is absent for the vortex flow, it is added for comparison with the case of turbulent boundary layer flow in a channel.}
\label{fig:shear}
\end{figure*}

Note that the differential relationship between the shear rate profile $\Sigma({r})$ and the velocity profile $U({r})$ can be rewritten in an integral form:
\begin{equation}\label{UviaSigma}
    U({r}) = -{r} \int\limits_{r}^R\frac{\Sigma({r}^\prime)}{{r}^\prime} \mathrm{d}{r}^\prime,
\end{equation}
where the distance $R$ plays a role of an integration constant. If $\Sigma({r})$ has permanent sign, then $U({r})$ has the opposite sign inside the tube of radius $R$, at ${r}<R$, and changes its sign outside at ${r}>R$. It is reasonable to assume that beyond the vortex the mean flow is absent or at least uncorrelated with the flow inside the vortex. Then the region ${r}\gtrsim R$ should be treated as the boundary of the vortex, where the axial symmetry is violated.

In what follows, we consider a statistically stationary situation, therefore $\partial_t U = 0$ and the moment of forces $-{r}^2 \Pi^{\varphi{r}}({r})$ acting on fluid in a circle of radius ${r}$ per unit length is also equal to zero. As a result, we arrive at the equation $\Pi^{\varphi{r}} = 0$ or
\begin{equation}\label{U}
    \langle u^{\varphi} u^{{r}} \rangle - \nu \Sigma = 0,
\end{equation}
which will further allow us to determine the velocity profile of a coherent vortex. It is important to note that equation (\ref{U}) requires the inequality
\begin{equation}\label{the-inequality}
    \Sigma \langle u^{\varphi} u^{{r}} \rangle >0
\end{equation}
to be satisfied throughout the vortex flow region.

To discuss the meaning of expression (\ref{the-inequality}), let us introduce the local Cartesian coordinate system $\{\xi,\eta,z\}$, whose origin moves around the axis of the vortex along a circle of radius $r$ with an angular velocity $U({r})/{r}$, see Fig.~\ref{fig:vortexr}. Its axes ${\bf e}_\xi$, ${\bf e}_\eta$ rotate with the same angular velocity. In this coordinate system, the directions of unit vectors ${\bf e}_{r}$, ${\bf e}_\varphi$ remain constant, and the distance to the vortex axis is given by $|{r}-\eta|$. The local assignment of the coordinate system means that we are only interested in distances $\eta\ll r$.

Consider, for definiteness, a vortex with $\Sigma(r)<0$. Then the distribution of the mean flow velocity near the origin of the introduced Cartesian coordinate system will be as shown in Fig.~\ref{fig:shear}. Inequality (\ref{the-inequality}) means that the turbulent part of the $\xi$-component of the momentum flux $\langle u^{\varphi} u^{{r}} \rangle$ must be directed to the vortex axis. This essentially distinguishes the situation in a vortex from, for example, a turbulent flow along a stationary boundary. Within the geometry of Fig.~\ref{fig:shear}, the plane $\eta = 0$ can correspond to the boundary, then the fluid flow will be directed along the $\xi$-axis. In this case, the turbulent momentum flux must be directed towards the boundary, see, e.g., Ref.~\onlinecite{smits2011high}, that is, in the opposite direction. In the case of a vortex, inequality (\ref{the-inequality}) is fulfilled due to the local rotation of the fluid $\Omega$, which is the sum of the external rotation $\Omega_0$ and the local addition $U({r})/{r}$,
\begin{equation}\label{localOmega}
    \Omega({r}) = \Omega_0 + U({r})/{r},
\end{equation}
see also Ref.~\onlinecite{balbus2017high}. If $U({r})$ has the same sign as $\Omega_0$, then the vortex is called a cyclone. According to relation (\ref{UviaSigma}), cyclone corresponds to different signs of $\Sigma$ and $\Omega$. An anticyclone is a vortex in which $U$ and $\Omega_0$ have opposite signs. The vortex is statistically stable if $\Omega(r)$ does not change its sign inside the vortex. Thus, the signs of $\Omega$ and $\Sigma$ coincide in an anticyclone.

\section{Energy balance}

\begin{figure*}
\begin{minipage}[ht]{0.49\linewidth}
\center{\includegraphics[width=1\linewidth]{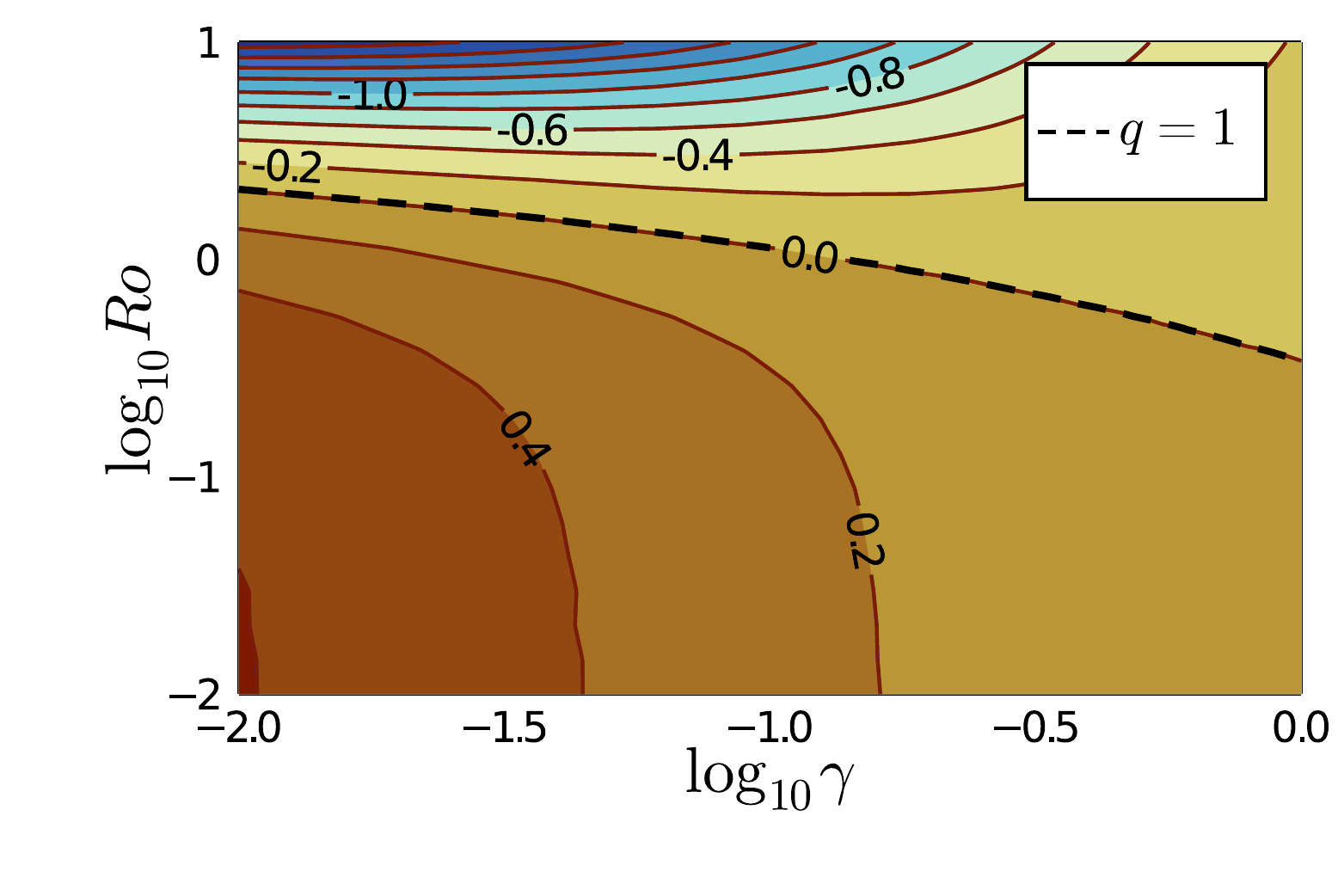}}
\end{minipage}
\hfill
\begin{minipage}[ht]{0.49\linewidth}
\center{\includegraphics[width=1\linewidth]{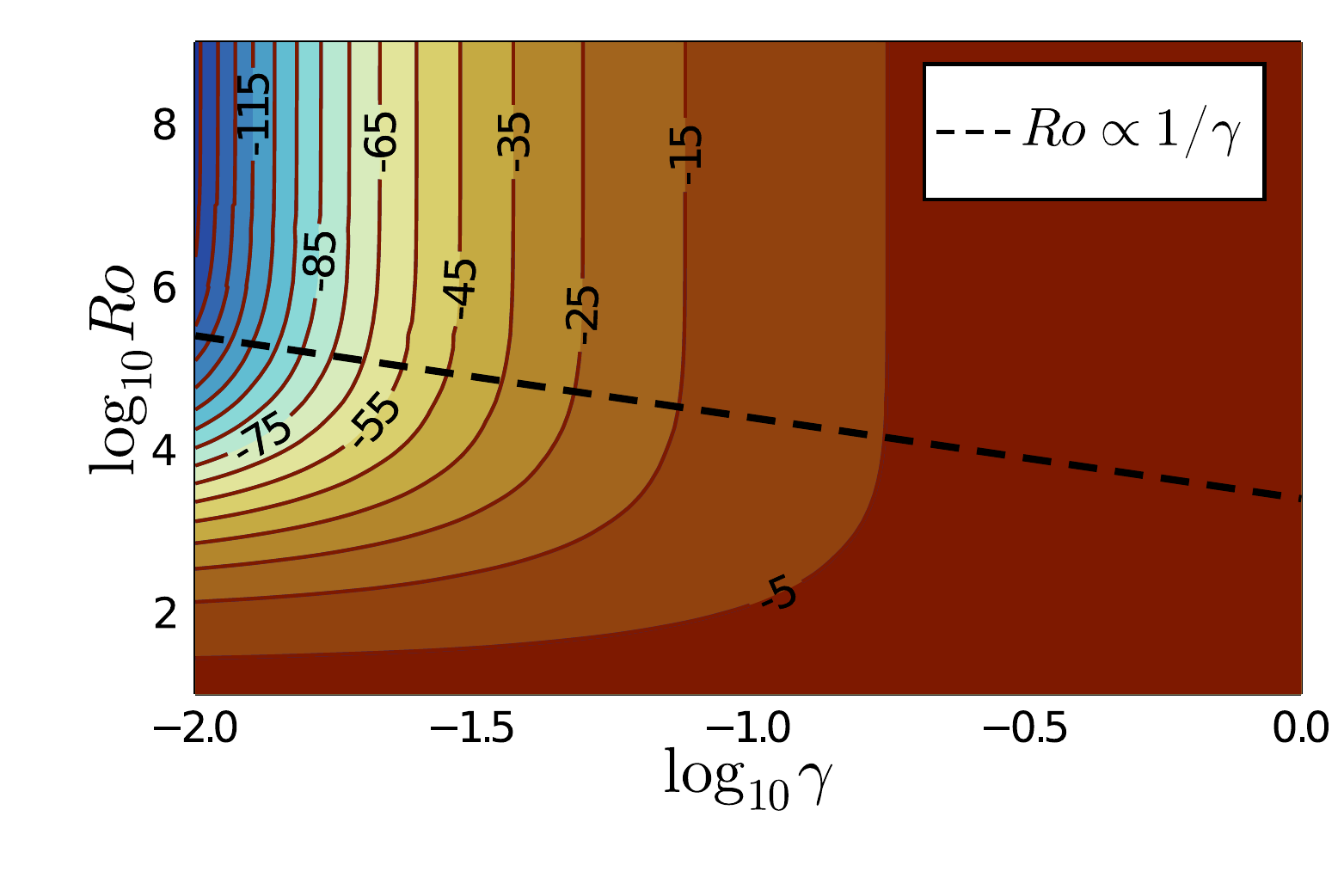}}
\end{minipage}
\caption{Numerical data for dimensionless tangent Reynolds stress $F(\mathrm{Ro},\gamma) = \Sigma \langle u^{\varphi} u^{{r}} \rangle/\epsilon$ at (a) small and (b) large positive Rossby numbers $\mathrm{Ro} = \Sigma/(2\Omega)>0$, which correspond to the anticyclone. In Fig.~\ref{fig:FpositiveRo}a the dashed line shows that the curve $q=1$ coincides with the curve $F=0$, see equation (\ref{epsilon-division}). In Fig.~\ref{fig:FpositiveRo}b the function $F(\mathrm{Ro},\gamma)$ ceases to depend on the parameter $\mathrm{Ro}$ above the dashed line following Ref.~\onlinecite{nazarenko2000nonlinear}.}
\label{fig:FpositiveRo}
\end{figure*}

The sign of the Reynolds stress tensor component $\langle u^{\varphi} u^{{r}} \rangle$ also determines the direction of the energy flow between the large-scale mean current ${\bm U}$ and small-scale turbulent pulsations. Let us analyze the energy balance based on the Navier-Stokes equation (\ref{ns_init}). To do this, we multiply this equation by $\bm{v}$ and carry out averaging over time:
\begin{equation}\label{energy_eq}
    \partial_t \langle {T}^{kin} \rangle
    +
    \mathop{\mathrm{div}}\langle{T}^{kin}\bm{v}\rangle
    =
    -\partial_i\langle v^i \, p \rangle + \nu \, \langle v^i \Delta v^i \rangle + \epsilon,
\end{equation}
where we have introduced the kinetic energy density ${T}^{kin} = \bm{v}^2/2$.

First, we consider the viscous term
\begin{equation}\label{eq:12}
\begin{aligned}
    \langle v^i \Delta v^i \rangle = U^i \Delta U^i + \langle u^i \Delta u^i \rangle = \\
    U(\partial_{{r}} + 2 / {r}) \Sigma + \frac{\Delta \langle {\bm u}^2\rangle}{2} - q \epsilon/ \nu,
\end{aligned}
\end{equation}
where $q = \nu\langle \partial_i u^k \partial_i u^k\rangle/\epsilon$ is the dimensionless rate of viscous dissipation of the kinetic energy of turbulent pulsations. We assume that the local approximation is valid, which means that the vortex size $R$ is much larger than the pumping scale $1/k_f$ and, therefore, the term $\Delta\langle {\bm u}^2\rangle /2$ in equation (\ref{eq:12}) is small compared to $q\epsilon/\nu$ as $1/(k_fR)^2$ and can be dropped.

Next, since we are analyzing the stationary regime, which implies the axial symmetry of the vortex, the kinetic energy flux is directed only along the radial direction, $\langle {T}^{kin}{\bm v}\rangle = \langle {T}^{kin}u^{r}\rangle{\bf e}_{r}$. The average amplitude is equal to
\begin{equation}\label{energy-flux}
    \langle {T}^{kin}u^{r}\rangle
    =
    U\langle u^{r} u^\varphi\rangle + \frac{1}{2}\langle {\bm u}^2 u^{r}\rangle.
\end{equation}
As turbulent pulsations are assumed to be weak, the last term in expression (\ref{energy-flux}) can also be neglected.

Finally, we analyze the term with pressure. Due to the axial symmetry of the vortex and its homogeneity along its axis, only the component $\langle v^r p\rangle = \langle u^r p\rangle$ should be considered. Next, we set $\langle u^r p\rangle=0$ following Ref.~\onlinecite{falkovich2016interaction}, where the absence of the pressure-velocity correlation was adopted as a hypothesis based on empirical observations \cite{laurie2014universal}. As a result, we arrive to energy balance equation for the statistically stable vortex flow, see below, which will be verified numerically in an independent way in Section~\ref{sec:pulsations} and thereby substantiate the validity of this assumption in our case.

Collecting all the results together and equating the rate of change in the average kinetic energy density to zero, we transform equation (\ref{energy_eq}) to the form
\begin{equation}\label{energy_eq_fin}
    \frac{1}{{r}} \, \partial_{{r}} \big({r} U \langle u^{{r}} u^{\varphi} \rangle\big)
    =
    \nu U(\partial_{{r}} + 2 / {r}) \Sigma + \epsilon(1-q).
\end{equation}
Now we use $\langle u^{{r}} u^{\varphi} \rangle = \nu \Sigma$ which follows from relation (\ref{U}), and obtain the final expression for the energy balance in the system
\begin{equation}\label{epsilon-division}
    \epsilon (1-q) =  \Sigma \langle u^{{r}} u^{\varphi} \rangle
    \qquad
    \mathrm{or} \qquad
    q = 1 - \frac{\Sigma}{\epsilon}\langle u^{r} u^\varphi\rangle.
\end{equation}
Thus, $\Sigma \langle u^{{r}} u^{\varphi} \rangle = \nu \Sigma^2$ is the power density that is transferred from turbulent pulsations to the large-scale coherent vortex. This power compensates for the viscous damping of the large-scale flow, and it should be positive throughout the vortex flow region, if only the coherent vortex exists, see equation (\ref{the-inequality}). Let us stress again that relation (\ref{epsilon-division}) takes place only if the vortex is large-scale, i.e. $k_f R \gg 1$.

Based on equation (\ref{epsilon-division}), one can estimate $\langle u^{r} u^\varphi\rangle\sim \epsilon/\Sigma$ and then, according to expression (\ref{U}), one finds $\Sigma \sim \sqrt{\epsilon/\nu}$. For further consideration, it is convenient to use these estimates to introduce the Rossby number for the large-scale flow $\mathrm{Ro}_{\scriptscriptstyle R} = \sqrt{\epsilon/\nu}/(2 \Omega_0)$. Together with the Reynolds number $\mathrm{Re}_f = \epsilon^{1/3}/\nu k_f^{4/3}$ characterizing the pumping, these two dimensionless parameters determine the behavior of the system. Note that if one takes the standard definition for small-scale Rossby number $\mathrm{Ro}_f = \epsilon^{1/3}k_f^{2/3}/2 \Omega_0$, then the large-scale Rossby number $\mathrm{Ro}_{\scriptscriptstyle R} = \mathrm{Ro}_f\sqrt{\mathrm{Re}_f}$.

\section{Statistics of turbulent pulsations}\label{sec:pulsations}

Equation (\ref{U}) contains a component $\langle u^{\varphi} u^{{r}} \rangle$ of the Reynolds stress tensor, which can be calculated if we assume that the gradient of the velocity ${\bm u}$ in the pulsations is small in comparison with the gradient of the mean current $\Sigma$. Then the evolution of velocity pulsations can be described in the linear approximation. In this approximation, equation (\ref{ns_init}) written in the Cartesian coordinate system $\{\xi,\eta,z\}$, which was introduced above, takes the form
\begin{equation}\label{u-dynamics-eq}
    (\partial_t-\Sigma\,\eta\,\partial_\xi)\bm u
    -
    u^\eta\,\Sigma\,{\bm e}_\xi+2\,\Omega\left[{\bm e}_z,\bm u\right]
    =
    -\nabla \tilde p+\nu\Delta\bm u+\bm f,
\end{equation}
where the local angular velocity $\Omega$ of fluid rotation was defined in expression (\ref{localOmega}). The random external force is assumed to be statistically isotropic and shortly correlated in time. Its pair correlation function is
\begin{equation}\label{force}
\begin{aligned}
    \langle f^i_{\bm{k}}(t_1) f^j_{\bm{q}}(t_2) \rangle = (2\pi)^3 \left(\delta^{ij} - \frac{k^ik^j}{{\bm k}^2}\right) \times \\ \epsilon \delta(t_1 - t_2) \delta(\bm{k} + \bm{q}) \chi(k)
\end{aligned}
\end{equation}
in the Fourier space. The spatial profile $\chi(k)$ tends to zero at $k\to 0$ and $k\to \infty$, so the force does not excite very large and very small eddies; it reaches maximum at $k\sim k_f$. For further calculations, we choose the Gaussian spatial profile
\begin{equation}\label{pumping-corr-fcn}
  \chi(k)=\dfrac{16\pi^{3/2}}{3k_f^5}{k}^2\exp\left[-\dfrac{{k}^2}{k_f^2}\right].
\end{equation}

\begin{figure}
\centering{\includegraphics[width=\linewidth]{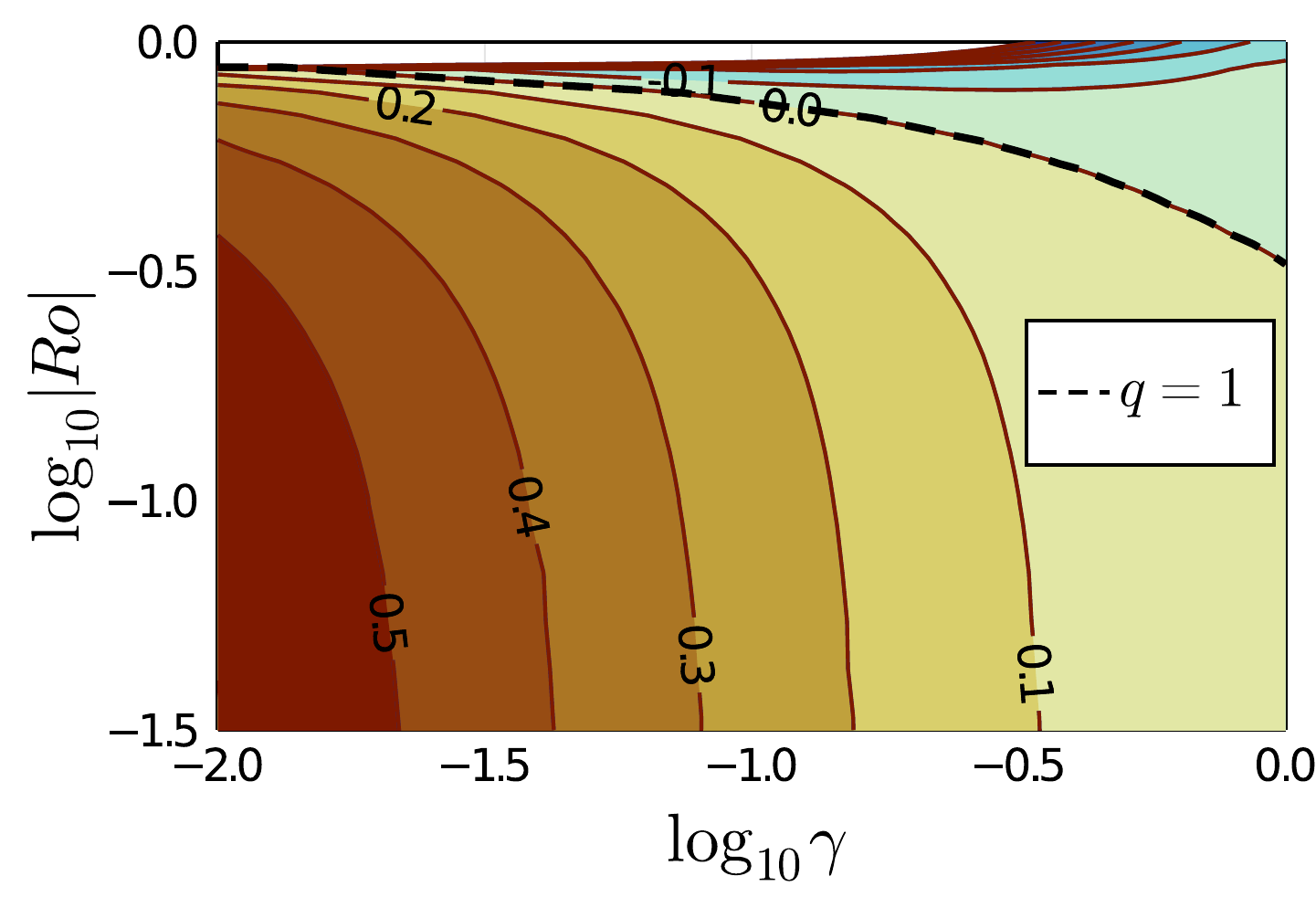}}
\caption{Numerical data for dimensionless tangent Reynolds stress $F(\mathrm{Ro},\gamma) = \Sigma \langle u^{\varphi} u^{{r}} \rangle/\epsilon$ at negative Rossby numbers $\mathrm{Ro} = \Sigma/(2\Omega)<0$, which correspond to the cyclone. The dashed line shows that the curve $q=1$ coincides with the curve $F=0$, see equation (\ref{epsilon-division}).}
\label{fig:FnegativeRo}
\end{figure}

The problem (\ref{u-dynamics-eq}) can be solved analytically \cite{salhi1997analysis} or numerically if the function ${\bm f}(t,{\bm r})$ is known and the parameters $\Omega$ and $\Sigma$ do not depend on time, see Appendix~\ref{app:dynamics}. The statistical properties of ${\bm u}$ are determined by two dimensionless parameters: the local Rossby number $\mathrm{Ro}$ for the large-scale flow and the local parameter $\gamma$ 
\begin{equation}\label{Ro-gamma}
    \mathrm{Ro} = \frac{\Sigma}{2\Omega},
    \qquad
    \gamma = \frac{2\nu k_f^2}{|\Sigma|}.
\end{equation}
${\mathrm{Ro}}$ is the local version of the global Rossby number ${\mathrm{Ro}}_{\scriptscriptstyle R}$. If one admits $\Sigma \sim \sqrt{\epsilon/\nu}$, then $\gamma\sim({\mathrm{Re}}_f)^{-2/3}\ll1$.

For the component $\langle u^{r} u^\varphi\rangle$ of the Reynolds stress tensor, we obtain
\begin{equation}\label{F-definition}
    \langle u^{r} u^\varphi\rangle
    =
    -
    \langle u^\eta u^\xi\rangle
    =
    \frac{\epsilon}{\Sigma} F(\mathrm{Ro},\gamma),
\end{equation}
where the function $F(\mathrm{Ro},\gamma)$ should be evaluated numerically based on equation (\ref{uiuj_final}) with $i=\eta$, $j=\xi$. The results are shown in Fig.~\ref{fig:FpositiveRo} and in Fig.~\ref{fig:FnegativeRo} for positive (anticyclone) and negative (cyclone) Rossby numbers $\mathrm{Ro}$ respectively. Note that in both cases $F$ has $\mathrm{Ro}$-independent limit at small Rossby numbers~\cite{kolokolov2020structure}.

To verify our hypothesis concerning the absence of correlation between pressure and velocity we independently compute the viscous dissipation rate $q$ of the kinetic energy of turbulent pulsations, see equation ($\ref{dissipation_int}$). According to energy balance (\ref{epsilon-division}), one finds
\begin{equation}\label{qF}
    q=1-F,
\end{equation}
where $F>0$ due to equation (\ref{the-inequality}). As both $q$ and $F$ are analytical functions of $\mathrm{Ro}$ and $\gamma$, equality (\ref{qF}) should also hold for negative values of $F$, although $F<0$ cannot correspond to the vortex interior. Our numerical calculations confirm equation (\ref{qF}). In particular, the dashed lines in Fig.~\ref{fig:FpositiveRo}a and in Fig.~\ref{fig:FnegativeRo} demonstrate that the curve $q=1$ coincides with the curve $F=0$. Also note, that the dimensionless function $F$ obeys the restriction $F<1$, since the viscous dissipation rate $q$ is always positive.

When Rossby number $\Ro$ and parameter $\gamma$ are small, the function $F$ approaches the unity, see Ref.~\onlinecite[Eq.(26)]{kolokolov2020structure}. One can find the asymptotic behavior
\begin{equation}\label{qF13}
    q = 1-F \approx q_{1/3} \gamma^{1/3},
\end{equation}
where the value of $q_{1/3} \approx 1.8$ was calculated numerically. Note that the value of the constant $q_{1/3}$ depends on the specific choice of the spatial pumping profile $\chi(k)$, see equation (\ref{pumping-corr-fcn}).

\section{Vortex velocity profile}

\begin{figure*}
\center{\includegraphics[width=\linewidth]{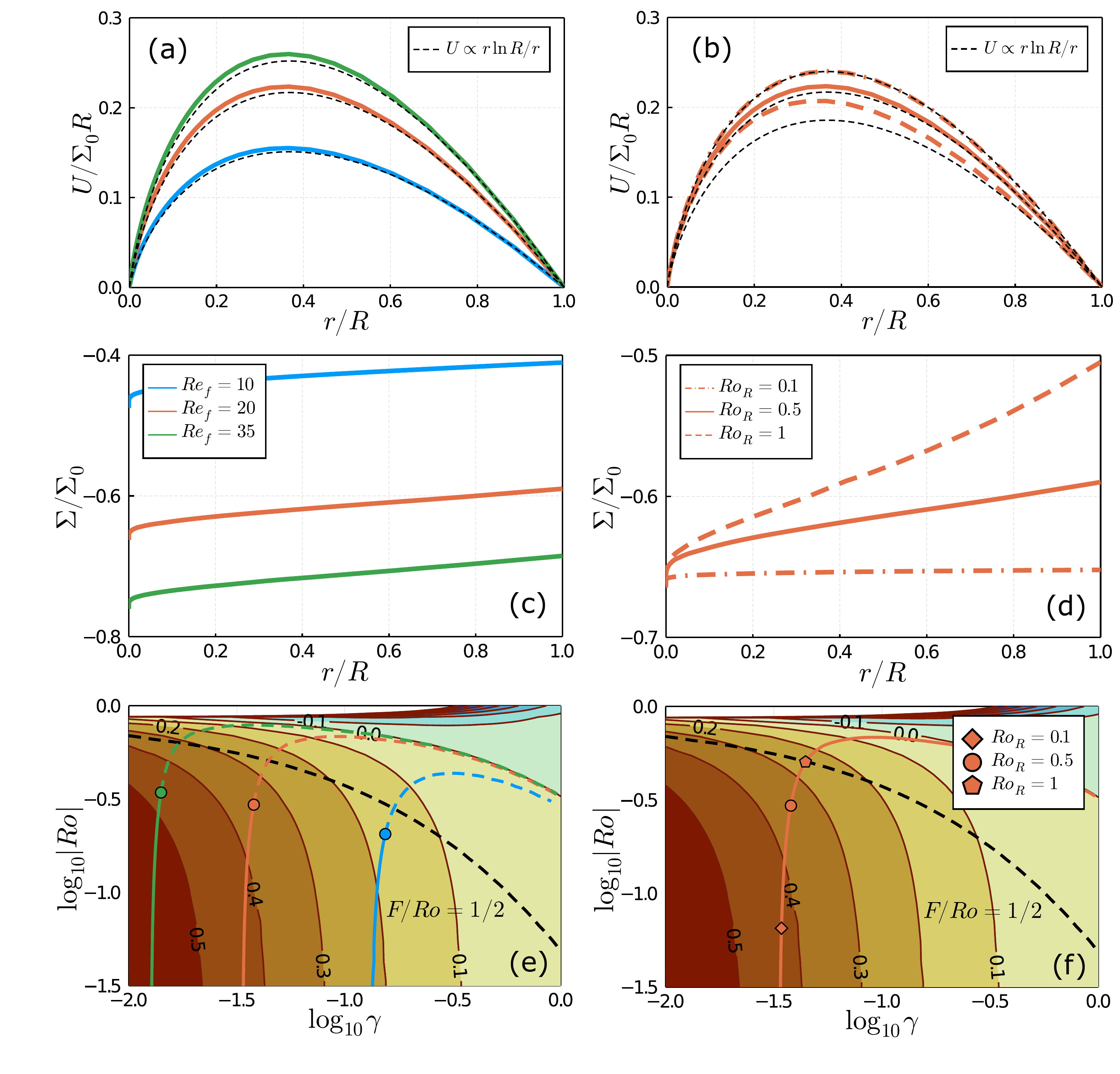}}
\caption{Dependence of the cyclone velocity profile on the parameters $\Re_f$ (left panel, $\Ro_R=0.5$) and $\Ro_R$ (right panel, $\Re_f=20$). (a,b) Azimuthal velocity of the cyclone $U$ and (c,d) parameter $\Sigma = r \partial_r (U/r)$ as a function of the distance $r/R$ to the vortex axis, $\Sigma_0=\sqrt{\epsilon/\nu}$. (e,f) Variation of the local parameters $\Ro$ and $\gamma$ inside the cyclone. The markers indicate the points corresponding to $r/R=1$. The area of curves downward from the marker corresponds to the movement to the vortex axis, in the opposite direction -- to its periphery (shown with dashed color lines in the left panel). The black dashed line corresponds to $F/\Ro=1/2$ (one-half takes into account the anisotropy of the velocity statistics) and above the line the linear analysis of turbulent pulsations becomes inapplicable.}
\label{fig:cyclone}
\end{figure*}

In this section we analyze the radial profile of the mean velocity $U(r)$ in the coherent vortex. First, let us rewrite equation (\ref{U}) in the dimensionless form. For this purpose, we introduce dimensionless variables: local shear rate $\sigma = \Sigma/\Sigma_0$, local rotation rate $\omega = 2 \Omega/\Sigma_0$ and the distance $\rho=r/R$ to the vortex axis, where $\Sigma_0=\sqrt{\epsilon/\nu}$. Then we obtain
\begin{eqnarray}\label{dimensionless-eq}
    \sigma^2
    =
    {F \left( \frac{\sigma}{\omega},
        \frac{2}{({\mathrm{Re}}_f)^{3/2} |\sigma|}\right)},
    \qquad
    \sigma = \rho \partial_\rho \omega/2,
\end{eqnarray}
and this essentially nonlinear first-order differential equation involving unknown function $\omega(\rho)$ should be supplemented by the boundary condition
\begin{equation}\label{eq:init}
    \omega\big\vert_{\rho=1}
    =
    \frac{1}{\mathrm{Ro}_{\scriptscriptstyle R}},
\end{equation}
which means that $U(R)=0$ according to expression (\ref{UviaSigma}). Note that equation (\ref{dimensionless-eq}) contains only one dimensionless parameter ${\mathrm{Re}}_f$, although the vortex structure generally depends on ${\mathrm{Re}}_f$ and $\mathrm{Ro}_{\scriptscriptstyle R}$. Thus, all solutions of equation (\ref{dimensionless-eq}) belong to a one-parameter family of curves, and the choice of a particular curve is determined by ${\mathrm{Re}}_f$. The value of $\mathrm{Ro}_{\scriptscriptstyle R}$ defines the position on the curve corresponding to the boundary or 'initial' condition (\ref{eq:init}), see Figs.~\ref{fig:cyclone},\ref{fig:anticyclone} and discussion below.

\subsection{Cyclones}

In the case of a cyclone, the negative root of equation~(\ref{dimensionless-eq}) should be chosen,
\begin{eqnarray}\label{cyclone-eq}
    \sigma
    =
    -
    \sqrt{F \left( \frac{\sigma}{\omega},
        \frac{2}{({\mathrm{Re}}_f)^{3/2}\cdot|\sigma|}\right)}, \quad
    \sigma = \rho \partial_\rho \omega/2.
\end{eqnarray}
We then numerically solve this differential-algebraic equation supplemented with the initial condition~($\ref{eq:init}$). The results are presented in Fig.~\ref{fig:cyclone}.

The left panel shows how the radial velocity profile depends on the forcing Reynolds number $Re_f$ at a fixed global Rossby number $\Ro_R=0.5$. Fig.~\ref{fig:cyclone}a demonstrates that with an increase in the Reynolds number, the velocity in the vortex increases, and the radial profile is quite close to the linear-logarithmic one. This corresponds to the almost complete independence of the local shear rate $\Sigma(r)$ on the distance $r$ to the vortex axis, see Fig.~\ref{fig:cyclone}c and equation ($\ref{UviaSigma}$). The dashed lines in Fig.~\ref{fig:cyclone}a show the dependence $U=-\Sigma^* r \ln R/r$, where $\Sigma^*=\Sigma(R)$. The angular velocity $U(r)/r$ associated with the self-rotation of the vortex increases when approaching the cyclone axis; therefore, the absolute value of the local Rossby number $\Ro$ decreases. This is illustrated in Fig.~\ref{fig:cyclone}e, where we show the phase portrait of equation~(\ref{cyclone-eq}) on the background of a color map corresponding to the values of $F(\Ro,\gamma)$ obtained in Section~\ref{sec:pulsations}. By changing the Reynolds number $\Re_f$, we move from one curve to another, and the marker corresponds to the coordinate $r=R$. The movement towards the axis of the cyclone in this diagram corresponds to the movement downward along the curves.

The right panel shows how the behavior of the system changes if we vary the global Rossby number $\Ro_R$ at a fixed value of the Reynolds number $\Re_f=20$. In this case, we remain on the same phase curve, but the position of the initial point $r=R$ changes, see Fig.~\ref{fig:cyclone}f. The larger the Rossby number $\Ro_R$, the higher we are on this curve and the more significant the difference between the radial velocity profile and the linear-logarithmic prediction, since the local parameter $\Sigma(r)$ begins to depend on $r$, see Figs.~\ref{fig:cyclone}b and \ref{fig:cyclone}d. Note that for the values $\Ro_R=1$ and $\Re_f=20$, the value $\Ro_f \approx 0.2 \lesssim 1$, which means that the fluid is still rotating rapidly.

In the limit ${\mathrm{Ro}}_{R}\ll1$ and $\Re_f \gg 1$ one can solve equation~(\ref{cyclone-eq}) analytically using the asymptotic expression~($\ref{qF13}$). The result is a linear-logarithmic velocity profile $U=-\tilde{\sigma} \Sigma_0  r \ln R/r$, where the parameter $\tilde \sigma$ is the minimum root of the equation
\begin{equation}\label{eq:ass}
  \tilde \sigma = - \sqrt{1-\dfrac{2^{1/3} q_{1/3}}{Re_f^{1/2} |\tilde{\sigma}|^{1/3}} }.
\end{equation}
In particular, for $\Re_f=20$, one finds $\tilde{\sigma} \approx - 0.64$, which is in reasonable agreement with the corresponding result presented in Fig.~\ref{fig:cyclone}d for $\Ro_R=0.1$.

The evolution of turbulent pulsations cannot be considered in the local approximation (adopted in Section~$\ref{sec:pulsations}$) near the axis of the vortex; therefore, our analysis should be limited to the region $r>r_c$, where $r_c$ defines the radius of the vortex core. Inside the core, inertial waves do not have time to transfer their energy to the large-scale flow before they leave the core. The group velocity of the waves is $v_g\sim \Omega(r_c)/k_f$ and the time needed for the energy transfer is $\sim 1/\Sigma(r_c)$. Thus, the core radius is determined by the equation \cite{kolokolov2020structure}
\begin{equation}\label{r_c}
    {\mathrm{Ro}}(r_c) \cdot k_f r_c \sim 1.
\end{equation}
Inside the core, the inertial waves are incoherently affected by inhomogeneous large-scale velocity field during their propagation in space and thus they do not produce considerable Reynolds stress $\langle u^r u^\varphi\rangle$. Therefore, one expects approximately rigid-body rotation of the fluid and the velocity profile has a linear dependence on $r$ in the region $r<r_c$.

Besides, the developed theory is not applicable in the region where the evolution of turbulent pulsations goes beyond the linear approximation. This occurs near the curve $F=0$, since in this case there is practically no energy transfer from fluctuations to the coherent vortex, see equation~(\ref{epsilon-division}). To establish the applicability condition of our consideration, we must compare the nonlinear interaction rate with the shear rate $|\Sigma|$. The nonlinear interaction rate $\Gamma_{\!nl}$ stems from the nonlinear term $({\bm u}\nabla){\bm u}$ omitted in equation (\ref{u-dynamics-eq}). Since the Rossby number is less than unity, the rate is diminished by the inertia wave oscillations and its upper estimation is $\Gamma_{\!nl}\lesssim \langle (\nabla {\bm u})^2\rangle /2\Omega$, see, e.g., Ref.~\onlinecite{galtier2003weak}. Some overestimation is due to the fact that the term $\langle (\nabla {\bm u})^2\rangle$ takes into account, inter alia, the spatial derivative normal to the ${\bm u}$-direction, whereas this derivative is absent in $({\bm u}\nabla){\bm u}$. The difference affects the result since the largest component is $u^\xi$ and the correlation length reaches a maximum in $\xi$-direction in the presence of a shear flow, see, e.g., Ref.~\onlinecite{champagne1970experiments}.  Next, in the linear approximation $\langle (\nabla {\bm u})^2\rangle \sim (1-F)\epsilon/\nu$ and using equation (\ref{anticyclone-eq00}), we can rewrite the inequality $\Gamma_{\!nl}\lesssim |\Sigma|$ as
\begin{equation}\label{uu-curveRe}
    F/\mathrm{Ro}\gtrsim1.
\end{equation}
The corresponding line is shown in Figs.~\ref{fig:cyclone}e and \ref{fig:cyclone}f (the factor $1/2$ instead of unity takes into account the aforementioned overestimation). 
Above this line, the applicability condition (\ref{uu-curveRe}) is violated, which means the development of three-dimensional turbulence. In this regime, the Reynolds stress tensor component $\langle u^r u^\varphi\rangle$ has the opposite sign, which makes the existence of coherent vortices impossible. Thus, the $\textquoteleft R{\text{\textquoteright}}$-point corresponding to the cyclone radius $r=R$ and indicated with the marker should be located below the black dashed curve in Figs.~\ref{fig:cyclone}e,f. This imposes restrictions on the global flow parameters ${\mathrm{Re}}_f,{\mathrm{Ro}}_{\scriptscriptstyle R}$ at which the cyclone can exist.

\subsection{Anticyclones}

\begin{figure*}
\center{\includegraphics[width=\linewidth]{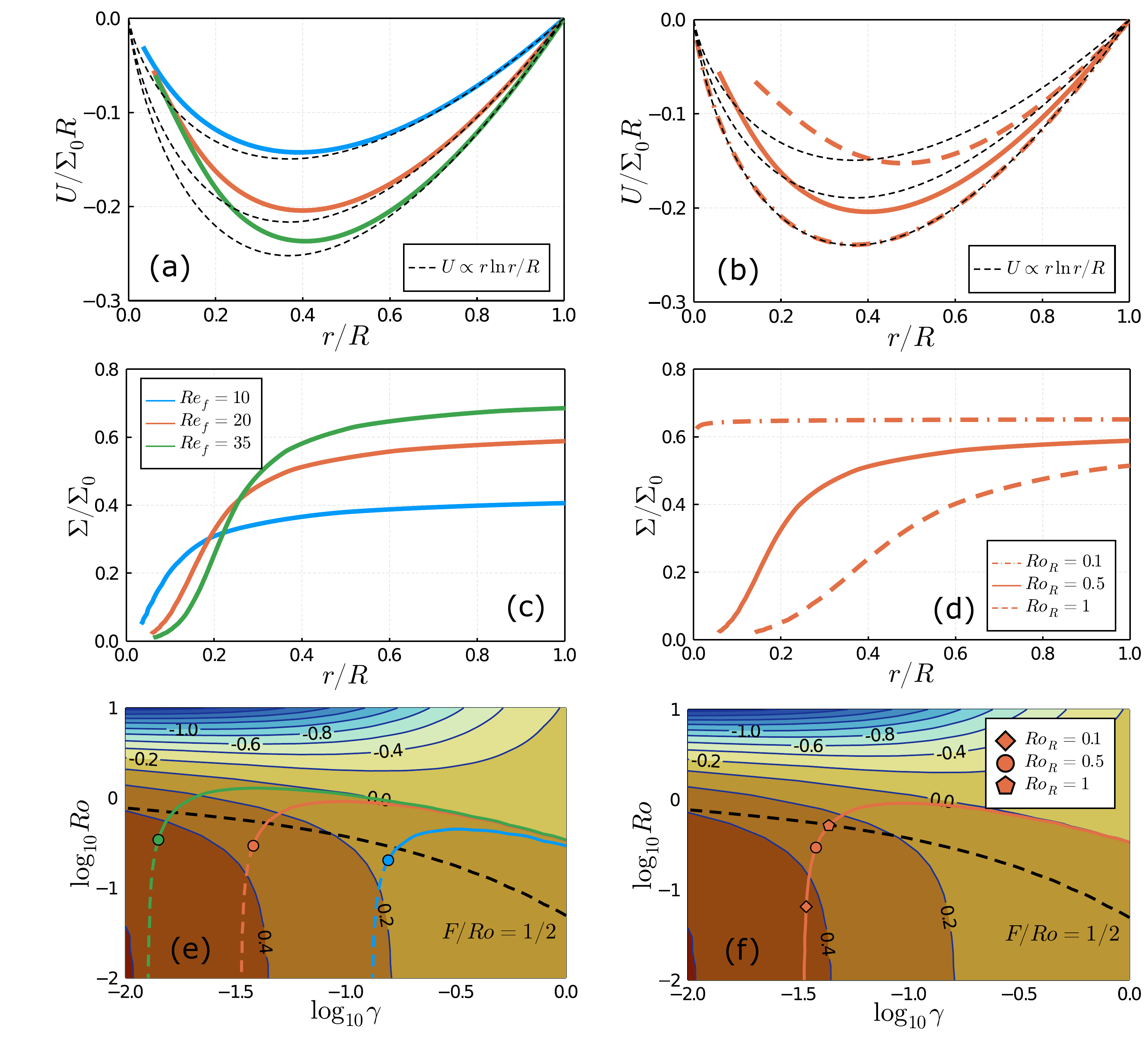}}
\caption{Dependence of the anticyclone velocity profile on the parameters $\Re_f$ (left panel, $\Ro_R=0.5$) and $\Ro_R$ (right panel, $\Re_f=20$). (a,b) Azimuthal velocity of the anticyclone $U$ and (c,d) parameter $\Sigma = r \partial_r (U/r)$ as a function of the distance $r/R$ to the vortex axis, $\Sigma_0=\sqrt{\epsilon/\nu}$. (e,f) Variation of the local parameters $\Ro$ and $\gamma$ inside the vortex. The markers indicate the points corresponding to $r/R=1$. The area of curves right-up from the marker corresponds to the movement to the vortex axis, in the opposite direction -- to its periphery (shown with dashed color lines in the left panel). The black dashed line corresponds to $F/\Ro=1/2$ (one-half takes into account the anisotropy of the velocity statistics) and above the line the linear analysis of turbulent pulsations becomes inapplicable.}
\label{fig:anticyclone}
\end{figure*}

In the case of an anticyclone, the positive root of equation~(\ref{dimensionless-eq}) should be chosen,
\begin{eqnarray}\label{anticyclone-eq00}
    \sigma
    =
    \sqrt{F \left( \frac{\sigma}{\omega},
        \frac{2}{({\mathrm{Re}}_f)^{3/2}\sigma}\right)}, \quad
    \sigma = \rho \partial_\rho \omega/2,
\end{eqnarray}
and a similar analysis of solutions to this equation is presented in Fig.~\ref{fig:anticyclone}. In comparison with cyclones under the same values of global parameters $\Re_f$ and $\Ro_R$, it should be noted that the maximum value of the velocity is lower in absolute value and the position of the maximum is farther from the axis of the vortex.

The self-rotation of the anticyclone $U(r)/r$ is directed opposite to the external rotation $\Omega_0$, and now the local Rossby number $\Ro$ increases when approaching the vortex axis. Accordingly, the movement along the phase curves shown in Figs.~\ref{fig:anticyclone}e and \ref{fig:anticyclone}f occurs in the opposite direction compared to cyclones. This leads to a stronger dependence of the local shear rate $\Sigma(r)$ on the distance to the vortex axis $r$ at the same values of the global parameters $\Re_f$ and $\Ro_R$, which is expressed in a noticeable difference between the velocity profile $U(r)$ and the linear-logarithmic prediction, see Figs.~\ref{fig:anticyclone}a-d. Good agreement is observed only for a small value of the Rossby number $\Ro_R=0.1$. In the limit $\Ro_R \ll 1$ and $\Re_f \gg 1$ the profiles of cyclones and anticyclones coincide up to sign, so the previous asymptotic expression ($\ref{eq:ass}$) can be used.

When moving along the phase curves, we inevitably come to the region $F \lesssim \Ro$, where the nonlinear interaction of fluctuations becomes essential, which should lead to the destruction of the anticyclone due to the direct energy cascade, see equation ($\ref{uu-curveRe}$). The only way to prevent this is to require that the size $r_c$ of the vortex core ($\ref{r_c}$) be large enough so that the point $r=r_c$ lies outside the region. As the shear rate $\Sigma$ vanishes inside the core, we should impose one more requirement to avoid the development of three-dimensional turbulence inside the core. It is suppressed if the rotation inside the core is large enough, ${\mathrm{Ro}}_\epsilon = (\epsilon k_f^2)^{1/3}/(2\Omega(r_c))<1$ (number ${\mathrm{Ro}}_\epsilon$ is the local version of ${\mathrm{Ro}}_f$). Then the external force ${\bm f}$ excites primarily inertia waves, which then spread away from the core. The requirement is always satisfied if the condition (\ref{uu-curveRe}) is met, since it can be rewritten in a form $F/(\gamma\cdot{\mathrm{Ro}}^3) > 1$ at $r=r_c$. Thus, it is necessary to take care only about the relation (\ref{uu-curveRe}), which should be fulfilled at $r=r_c$.

Next, it follows from equation (\ref{anticyclone-eq00}) that the ratio between the vortex and the core radii is
\begin{equation}\label{Rr_c}
    \frac{R}{r_c}
    =
        \exp\left(
        \int_R^{r_c} \frac{\mathrm{d}\ln(\gamma\cdot{\mathrm{Ro}})}
            {2{\mathrm{Ro}}}
        \right),
\end{equation}
where the integral should be taken along the phase curve. The location of $\textquoteleft \!R\text{\textquoteright}$-point on the phase plane is determined by the global parameters ${\mathrm{Ro}}_{\scriptscriptstyle R}$ and ${\mathrm{Re}}_f$, see the marked points in Figs.~\ref{fig:anticyclone}e,f. Now suppose for simplicity that $\textquoteleft r_c\text{\textquoteright}$-point corresponds to $\mathrm{Ro}(r_c)\ll1$. Then the parameter $\gamma$ remains approximately constant along the phase curve and we can perform integration in equation (\ref{Rr_c})
\begin{equation}
    2\ln\frac{R}{r_c}
    =
    \frac{1}{{\mathrm{Ro}}(R)}
    -
    \frac{1}{{\mathrm{Ro}}(r_c)}
    <
    \frac{1}{{\mathrm{Ro}}(R)}.
\end{equation}
Condition (\ref{eq:init}) implies $\mathrm{Ro}(R)=\sigma(R)\cdot \Ro_R$, and in the limit $\Ro_R \ll 1$ and $\Re_f \gg 1$ one finds $\sigma(R)=1$, see Ref.~\onlinecite{kolokolov2020structure}. The criterion in this form was mentioned in the Introduction.

In the general case, one needs to find numerically the value of $\rho_{nl} = r_{nl}/R$ that corresponds to the intersection between the black dashed line (\ref{uu-curveRe}) and the phase curve shown in Figs.~\ref{fig:anticyclone}e,f. Note that this value will depend only on the global flow parameters $\Re_f$ and $\Ro_R$, i.e. $\rho_{nl}=\rho_{nl}(\Re_f,\Ro_R)$. Next, as we discussed above, $r_c \geq r_{nl}$ and using equation (\ref{r_c}) we arrive at
\begin{equation}\label{eq:lim}
 R < R_{max} = \dfrac{1}{k_f} \max_{\rho_{nl}\leq \rho \leq 1} \dfrac{1}{\rho \Ro(\rho)}.
\end{equation}
The value of $R_{max}$ should be calculated numerically for specific values of $\Ro_R$ and $\Re_f$. For the parameters under study, the maximum is reached at point $\rho=\rho_{nl}$, which corresponds to $r_c=r_{nl}$. In particular, for $\Re_f=20$ and $\Ro_R=0.5$ one finds $k_f R_{max} \sim 5$ and for $\Re_f=20$ and $\Ro_R=0.1$ we obtain $k_f R_{max} \sim 2.6 \times 10^3$. If the size of the system $L$ substantially exceeds $R_{max}$ then we expect a difference between the sizes of cyclones and anticyclones in the condensate, since the size of cyclones is limited only by the system size.

\section{Concluding remarks}

To summarize, we considered a coherent vortex in a condensate of turbulent flow in a rotating fluid when Rossby number $\Ro_R \lesssim 1$ is small and Reynolds number $\Re_f \gtrsim 10$ is large. In the limit $\Ro_R \ll 1$, the radial profiles of the mean azimuthal velocity in cyclones and anticyclones are equal to each other up to a sign, $U(r) = \mp \tilde{\sigma} \Sigma_0 r \ln R/r$, where $R$ is the vortex radius, $\Sigma_0 = \sqrt{\epsilon/\nu}$, and $\tilde{\sigma}$ is determined by equation (\ref{eq:ass}). At finite Rossby numbers, we found some differences between them. Qualitatively, the maximum of $|U(r)|$ shifts to the center inside cyclones and to the periphery inside anticyclones, and the maximum value of $|U(r)|$ is greater inside cyclones than inside anticyclones at the same global parameters of the flow.

Coherent vortices can exist only if condition (\ref{uu-curveRe}) is met at the point $r=R$, which imposes restrictions on the allowable range of $\Re_f$ and $\Ro_R$. The induced rotation $U(r)/r$ of the cyclone is added to the external rotation $\Omega_0$, so the fulfillment of inequality (\ref{uu-curveRe}) at a point $r=R$ guarantees that it will be satisfied at any other point $r<R$ inside the vortex. In the case of anticyclones, the induced rotation compensates the external rotation, so the movement along the phase curves in Figs.~\ref{fig:anticyclone}e,f occurs in the opposite direction. This imposes an additional condition on the existence of anticyclones -- inequality (\ref{uu-curveRe}) must be fulfilled on the boundary of the vortex core, which size $r_c$ is mainly determined by the forcing scale, see equation (\ref{r_c}). This condition leads to the fact that the size $R$ of the anticyclone is limited by $R_{max}$ that depends on the global flow parameters $\Re_f$ and $\Ro_R$, see equation (\ref{eq:lim}). We predict that if the size of the system $L$ significantly exceeds $R_{max}$, then the sizes of cyclones and anticyclones in the condensate may also differ significantly. One should remember that expression (\ref{uu-curveRe}) is only an estimate, and the exact criterion must be obtained either by numerical simulation or experimentally. Besides, our theory is applicable only if the local approximation is valid for describing small-scale turbulent pulsations, when ${\mathrm{Ro}}_{\scriptscriptstyle R}\cdot k_f L\gg 1$.

As for comparing the developed theory with experiments, we should emphasize that our approach takes into account only bulk viscous dissipation without any influence of boundaries. In experiments, the influence of the lower and upper boundaries can be significant. Indeed, the thickness of the boundary layer is $\delta\sim\sqrt{\nu/\Omega_0}$, and then the scale $r_\alpha$ at which the bottom friction becomes comparable with the bulk viscosity is $r_\alpha \sim (\nu/\Omega_0)^{1/4} H^{1/2}$, where $H$ is the height of the vessel. For typical values $H=50\,\text{cm}$, $\nu=0.01\,\text{cm}^2/\text{s}$, $\Omega_0=3\,\text{s}^{-1}$, see Refs.~\onlinecite{ruppert2005extraction,boffetta2020cyclone,godeferd2015structure}, one finds $r_\alpha\sim 2\,\text{cm}$ that is of the order or smaller than the radii of the observed columnar vortices. Thus, the predictions of our theory can be directly applied only to the results of numerical simulations with periodic boundary conditions, while the description of the experiments requires further research.

\acknowledgments

We are grateful to V. Lebedev and I. Kolokolov for valuable discussions. This work was supported by the Russian Science Foundation, Grant No. 20-12-00383. SV and IV were supported by Grant No.
19-1-2-46-1 of Foundation for the Advancement of Theoretical Physics and Mathematics \textquotedblleft BASIS.\textquotedblright

\section*{Data AVAILABILITY}

The data that support the findings of this study are available from the corresponding author upon reasonable request.

\appendix

\begin{widetext}

\section{Dynamics and statistical properties of turbulent pulsations}\label{app:dynamics}

It is convenient to solve equation (\ref{u-dynamics-eq}) in the Fourier space
\begin{equation}\label{ns_u_init_fourier}
    \left(\partial_t+\Sigma\,k_{\xi}\,\partial_{k_{\eta}}\right)\bm u_{\bm k}
    -
    u^{\eta}_{\bm k}\,\Sigma\,{\bm e}_{\xi}
    +
    2\,\Omega\left[{\bm e}_z,\bm u_{\bm k}\right]
    =
    -i\bm k\tilde p_{\bm k}
    -
    \nu{\bm k}^2\bm u_{\bm k}+\bm f_{\bm k}
\end{equation}
using the method of characteristics. Then one has to solve the ordinary differential equation
\begin{equation}\label{ns_u_init_fourier_char}
    \dot{\bm u}_{\bm K}
    =
    u^{\eta}_{\bm K}\,\Sigma\,{\bm e}_{\xi}
    -
    2\,\Omega\left[{\bm e}_z,\bm u_{\bm K}\right]
    -
    i\bm K\tilde p_{\bm K}-\nu{\bm K}^2\bm u_{\bm K}
    +
    \bm f_{\bm K}
\end{equation}
along the characteristics
\begin{equation}\label{char}
    \bm K(t)=\{k_{\xi},\,K_{\eta}(t),\,k_z\}, \quad K_{\eta}(t)=k_{\eta}+\Sigma\,t\,k_{\xi}.
\end{equation}
To exclude the pressure term, we note that $   \mathrm{d}\left(\bm K,\,{\bm u}_{\bm K}\right)/\mathrm{d}t = 0$, so the projection of (\ref{ns_u_init_fourier_char}) onto ${\bm K}$ gives
\begin{equation}\label{pres_k}
    \tilde p_{\bm K}
    =
    \frac{2i}{{\bm K}^2}\biggl(\Omega k_{\eta} u^{\xi}_{\bm K}-\left(\Omega+\Sigma\right)k_{\xi}u^{\eta}_{\bm K}\biggr).
\end{equation}

Next, we substitute expression (\ref{pres_k}) into equation (\ref{ns_u_init_fourier_char}) and rewrite it in a matrix form
\begin{equation}\label{ns_u_mat_C}
    \dot{\bm u}_{\bm K}=\hat{C}\bm u_{\bm K}-\nu{\bm K}^2\bm u_{\bm K}+\bm f_{\bm K}, \qquad
    \hat{C}=
    \left(\begin{array}{ccc}
        2\,\Omega\frac{k_{\xi}K_{\eta}}{K^2} &
            \Sigma\left(1-2\frac{k_{\xi}^2}{K^2}\right)+2\,\Omega\left(1-\frac{k_{\xi}^2}{K^2}\right)
            &
            0
        \\
        -2\,\Omega\left(1-\frac{K_{\eta}^2}{K^2}\right) &
            -(\Omega+\Sigma)\frac{2k_{\xi}K_{\eta}}{K^2} &
            0
        \\
        2\,\Omega\frac{K_{\eta}k_z}{K^2} &
            -(\Omega+\Sigma)\frac{2k_{\xi}k_z}{K^2} &
            0
    \end{array}\right),
\end{equation}
where ${\bm u}_{\bm K}$ is assumed to be a column vector. Due to the incompressibility condition, the velocity vector ${\bm u}_{\bm K}$ depends only on two independent scalar functions. In order to derive the equations on these two functions, we follow Ref.~\cite{salhi1997analysis} and we choose basis $\left\{\bm{e}_1,\,\bm{e}_2,\,\bm{e}_3\right\}$
\begin{equation}\label{basis}
    \bm{e}_3=\frac{\bm{K}}{K},
    \quad
    \bm{e}_1=\frac{\left[{\bf e}_\eta,\bm{e}_3\right]}
        {\left|\left[   {\bf e}_\eta,\bm{e}_3\right]\right|},
    \quad
    \bm{e}_2=\left[\bm{e}_3,\bm{e}_1\right].
\end{equation}
The expansions of vectors $\bm{u}_{\bm{K}}$ and $\bm{f}_{\bm{K}}$ in this basis are
\begin{equation}\label{psi}
    \bm{u}_{\bm{K}}
    =
    \sum_\alpha \psi_{\bm{K}}^\alpha\bm{e}_\alpha,
    \quad
    \bm{f}_{\bm{K}}
    =
    \sum_\alpha
    \phi_{\bm{K}}^\alpha\bm{e}_\alpha,
    \qquad
    \alpha,\beta=\{1,2\},
\end{equation}
but one should remember that the basis (\ref{basis}) is time dependent when substituting the expansion (\ref{psi}) into the evolution equation (\ref{ns_u_mat_C}). The result of the corresponding substitution is
\begin{equation}\label{equat}
    \frac{d{}}{dt}
    \Psi_{\bm{K}}
    =
    \hat{H}\Psi_{\bm{K}}
    -\nu\mathbf{K}^2
    \Psi_{\bm{K}}
    +
    \Phi_{\bm{K}},
    \quad
    \Psi_{\bm{K}}
    =
    \begin{pmatrix}
    \psi^1_{\bm K} \\
    \psi^2_{\bm K} \\
    \end{pmatrix},
    \quad
    \Phi_{\bm{K}}
    =
    \begin{pmatrix}
    \phi^1_{\bm K} \\
    \phi^2_{\bm K} \\
    \end{pmatrix},
    \quad
    \hat H =
    \frac{1}{K^2}\begin{pmatrix}
    0 & (\Sigma+2\,\Omega)Kk_z\\\\
    -2\,\Omega K k_z & -\Sigma k_{\xi}K_{\eta}
    \end{pmatrix}.
\end{equation}
The pair correlation function of the external force (\ref{force}) in the basis $\left\{\bm{e}_1,\,\bm{e}_2,\,\bm{e}_3\right\}$ is equal to
\begin{equation}\label{forceapp}
    \langle \phi^\alpha_{\bm k}(t)\,\phi^\beta_{\bm q}(t^\prime)\rangle
    =
    (2\pi)^3 \delta({\bm k}+{\bm q})
    \big(\delta^{\alpha1}\delta^{\beta1} - \delta^{\alpha2}\delta^{\beta2}\big)
    \epsilon \delta(t - t^\prime) \chi(k).
\end{equation}
The alternating sign in the tensor structure of equation (\ref{forceapp}) reflects the fact that the basis (\ref{basis}) changes its orientation when ${\bm k}\to-{\bm k}$.

The solution of equation~(\ref{equat}) can be represented in the form
\begin{equation}
    \Psi_{\bm{K}(t)}(t)
    =
    \int\limits_{-\infty}^{t}\mathrm{d}t^\prime
    \exp\left[-\nu\int\limits_{t^\prime}^t
        \mathrm{d} t^{\prime\prime}\,
        \bm{K}^2(t^{\prime\prime})\right]
    \hat{Q}(t,\,t^{\prime})\,
    \Phi_{\bm{K}(t^{\prime})}(t^{\prime}),
\label{solution_equat}
\end{equation}
where the evolution matrix $\hat{Q}(t,t^{\prime})$ obeys the equation
\begin{equation}\label{evolution_matrix_problem}
    \frac{d{}}{dt}\,\hat{Q}(t,t^{\prime})
    =
    \hat{H}(t)\hat{Q}(t,t^{\prime}),
    \qquad
    \hat{Q}(t^{\prime},t^{\prime})=\hat{1}.
\end{equation}
However, for numerical calculations it will be more convenient to set the final moment and then the evolution matrix $\hat Q$ also obeys the backward equation
\begin{equation}\label{evolution_matrix_problem_tau}
    \frac{d}{dt^{\prime}}\,\hat{Q}(t,t^{\prime}) = -\hat{Q}(t,t^{\prime})\,
    \hat{H}(t^{\prime}),\qquad
    \hat{Q}(t,\,t) = \hat{1}.
\end{equation}
Let us note that the matrix $\hat H$ (\ref{equat}) depends only on the direction of wavevector ${\bm k}$, but not on its absolute value. Therefore, the same is true for the evolution matrix $\hat Q$.

Now we are ready to proceed to the direct calculation of the Reynolds stress tensor. According to the introduced definitions,
\begin{equation}
\langle u^i(t)u^j(t)\rangle=\int\dfrac{d^3\bm{k}}{(2\pi)^3}\int\dfrac{d^3\bm{s}}{(2\pi)^3}\langle u^i_{\bm{K}}(t)u^j_{\bm{S}}(t)\rangle, \quad
\langle u^i_{\bm{K}}(t)u^j_{\bm{S}}(t)\rangle=\sum_{\alpha,\beta}\,\langle \psi^{\alpha}_{\bm{K}}(t)\psi^{\beta}_{\bm{S}}(t)\rangle e^i_{\alpha}(\bm{K}(t))e^j_{\beta}(\bm{S}(t)),
\label{u_correl_1}
\end{equation}
and using expression \eqref{solution_equat} we obtain
\begin{equation}\label{psi_correl_1}
    \begin{aligned}
    \langle \psi^{\alpha}_{\bm{K}}(t)\psi^{\beta}_{\bm{S}}(t)\rangle
    =
    \int\limits_{-\infty}^{t}dt_1 dt_2
    \exp\left[-\nu\left(
        \int\limits_{t_1}^{t}dt'_1\,\bm{K}^2(t'_1)
        +
        \int\limits_{t_2}^{t}dt'_2\,\bm{S}^2(t'_2)
        \right)\right] \times\\
    \sum_{\sigma,\rho}\,
    Q^{\alpha\sigma}\bigl(\bm{K}(t),\bm{K}(t_1)\bigr)
    Q^{\beta\rho}\bigl(\bm{S}(t),\bm{S}(t_2)\bigr)
    \langle \phi^{\sigma}_{\bm{K}}(t_1)\phi^{\rho}_{\bm{S}}(t_2)\rangle.
    \end{aligned}
\end{equation}
Since the statistics of the turbulent flow is assumed to be stationary in time, we can set $t=0$ in equation (\ref{psi_correl_1}). Further calculations, taking into account relation (\ref{forceapp}), are rather straightforward and we omit them for brevity. We present the final results using the introduced parameters (\ref{Ro-gamma}). Besides, we also introduce the dimensionless time $\tau = -|\Sigma| t^\prime>0$, where $t^\prime$ appears in equation (\ref{evolution_matrix_problem_tau}), and $\varsigma =  - \mathop{\mathrm{sign}}\Sigma$, so that $\varsigma=1$ and $\varsigma=-1$ correspond to cyclone and anticyclone respectively. We also define
\begin{equation}
\begin{aligned}
\lambda(\theta, \, \varphi, \, \tau, \, \varsigma)=1+\varsigma\tau\sin2\varphi\sin^2\theta+(\tau\cos\varphi\sin\theta)^2, \qquad
\Lambda(\theta, \, \varphi, \, \tau, \, \varsigma)=\int\limits_0^{\tau} d\tau' \lambda(\theta, \, \varphi, \, \tau', \, \varsigma),
\end{aligned}
\label{lambda_sigma}
\end{equation}
and then the Reynolds stress tensor components are equal to
\begin{equation}\label{uiuj_final}
   \langle u^i u^j \rangle
    =
    \dfrac{\epsilon}{\pi |\Sigma|} \,
    \int \limits_0^{\pi/2} d\theta
    \int \limits_0^{\pi} d\varphi
    \int \limits_0^{+\infty} d\tau \,
    \frac{\lambda \, \sin\theta}
        {\left(\lambda + \gamma \Lambda\right)^{-5/2}}
    \sum_{\alpha,\beta,\delta = 1,2}
        e_{\alpha}^i\,e_{\beta}^j\,
        Q^{\alpha \delta}\,Q^{\beta \delta},
\end{equation}
where we used the Gaussian spatial profile ($\ref{pumping-corr-fcn}$) for the pumping. In a similar way, for the dimensionless viscous dissipation rate we obtain
\begin{equation}\label{dissipation_int}
    q
    =
    \dfrac{5}{2\pi}
    \int \limits_0^{\pi/2} d\theta
    \int \limits_0^{\pi} d\varphi
    \int \limits_0^{+\infty} d\tau \,
    \frac{\sin\theta}
        {\left(\lambda + \gamma \Lambda\right)^{-7/2}}
    \sum_{\alpha,\beta = 1,2}
        \big(Q^{\alpha \beta}\big)^2.
\end{equation}
The evolution matrix $Q^{\alpha \beta}$ should be found by solving equation (\ref{evolution_matrix_problem_tau}), which takes the form
\begin{equation}\label{evolution_matrix_problem_final}
    \dfrac{\partial{}}{\partial{\tau}}\,\hat{Q}(\tau)
    =
    \hat{Q}(\tau)
    \begin{pmatrix}
    0 &
    -\varsigma\left(1 + \dfrac{1}{\mathrm{Ro}}\right)
        \dfrac{\cos\theta}{\sqrt{\lambda(\tau)}}
    \\[5pt]
    \dfrac{\varsigma}{\mathrm{Ro}}
        \dfrac{\cos\theta}{\sqrt{\lambda(\tau)}}
    & \dfrac{\cos\varphi\sin^2\theta
            (\tau\cos\varphi
            +
            \varsigma\sin\varphi)}
        {\lambda(\tau)}
    \end{pmatrix}, \qquad \hat{Q}(0) = \hat{1}.
\end{equation}


\end{widetext}

\bibliography{voro}

\end{document}